\documentclass[aps,prl,twocolumn,superscriptaddress,showpacs]{revtex4}
\usepackage{graphicx}
\usepackage{graphicx,epsfig}
\usepackage{amssymb,fancyheadings,times}
\usepackage{amsmath,amsfonts,bm}
\newcommand \bea {\begin{eqnarray}}
\newcommand \eea {\end{eqnarray}}
\newcommand \ba {\begin{eqnarray*}}
\newcommand \ea {\end{eqnarray*}}
\newcommand \be {\begin{equation}}
\newcommand \ee {\end{equation}}
\newcommand \dagga {{\phantom{\dagger}}}

\begin{document}
\title{Surprises in the phase diagram of an Anderson impurity model for a
single C$_{60}^{n-}$ molecule}
\author{Lorenzo De Leo}
\affiliation{Department of Physics and Center for Material Theory,
Rutgers University, Piscataway, NJ 08854, USA}
\author{Michele Fabrizio}
\affiliation{International School for Advanced Studies (SISSA),
%and Instituto
%Nazionale per la Fisica della Materia (INFM) UR-Trieste SISSA,
Via Beirut 2-4,
I-34014 Trieste, Italy}
\affiliation{The Abdus Salam International Center for Theoretical Physics
(ICTP),
P.O.Box 586, I-34014 Trieste, Italy}
\begin{abstract}
We find by Wilson numerical renormalization group and conformal
field theory that a three-orbital Anderson impurity model for a
C$_{60}^{n-}$ molecule has a very rich phase diagram which
includes non-Fermi-liquid stable and unstable fixed points with
interesting properties, most notably high sensitivity to doping
$n$. We discuss the implications of our results to the conductance
behavior of C$_{60}$-based single-molecule transistor devices.
\end{abstract}
\maketitle
The characteristic behavior of an Anderson impurity model (AIM) emerges often unexpectedly
in physical contexts which are apparently faraway from magnetic alloys.
The typical example is the conductance behavior of nano-scale devices, {\sl e.g.}
quantum dots and single-molecule transistors (SMT), where Kondo-assisted tunneling
may lead to nearly
perfect transmission at zero bias in spite of a large charging energy\cite{QD-theory}.
Usually the two conduction leads are bridged by a single quantum dot or molecular level, which
therefore behaves effectively as a one-orbital AIM. Less common is the case when several levels
happen to be nearby degenerate thus realizing in practice a multi-orbital AIM.
This may be achieved for instance in SMTs built with high-symmetry molecules.
Recently a zero-bias anomaly has been reported for a C$_{60}$ based SMT\cite{fullerene-SMT},
which has been proven to be of the Kondo-type by its splitting under the action of a magnetic
field or magnetic leads\cite{fullerene-SMT-Kondo}.
Fullerene's large electron affinity is known to yield to
electron transfer into C$_{60}$ when adsorbed on metallic substrates.
The actual number of doped electrons depends on the substrate\cite{Modesti}, but may also
be controlled by attaching alkali atoms to the molecule\cite{alkali}. In this case
a variety of conductance behavior has been reported depending on the number of
K-atoms attached to C$_{60}$\cite{alkali}, ranging from Kondo-like resonances to Fano-like
anti-resonances.

Motivated by this opportunity, in this Letter we study
the behavior of an AIM for C$_{60}^{n-}$ by
Wilson Numerical Renormalization Group (NRG)\cite{NRG}
and Conformal Field Theory (CFT)\cite{CFT}. We obtain a phase diagram which includes
Fermi- and non-Fermi-liquid phases, with a doping dependence
qualitatively in agreement with experiments.
In particular we find a Kondo-like zero-bias anomaly for doping $n=1$,
likely the case of C$_{60}$ on Au\cite{Modesti}, as found in
Refs.\cite{fullerene-SMT,fullerene-SMT-Kondo}. For $n=2$, which should correspond to
pure or slightly K-doped C$_{60}$ on Ag\cite{Modesti},
we predict instead a conductance-minimum at zero-bias,
compatible with actual observations\cite{alkali}. Finally, for $n=3$ we
find a conductance minimum with a non-analytic voltage behavior $G(V)-G(0)\sim |V|^{2/5}$.

The LUMO's of C$_{60}$ are three-fold degenerate t$_{1u}$
orbitals\cite{Gunnarsson}. The electron-electron interaction acts
as if these orbitals were effectively $p$-orbitals. Therefore the
Coulomb energy of a C$_{60}^{n-}$ molecule with $n$ valence
electrons into a configuration with total spin $S$ and angular
momentum $L$ is ${\cal{H}}_{int} = {\cal{H}}_U + {\cal{H}}_{J}$,
with ${\cal{H}}_U = U\,\left(n-n_0\right)^2/2$ and ${\cal{H}}_{J}
= -J\left[2S(S+1)+\frac{1}{2}L(L+1)\right]$. The reference
valency, $n_0$, is controlled in SMTs by the gate voltage, the
metal substrate and the alkali doping. The Coulomb exchange,
$J>0$, favoring high-degeneracy states (conventional Hund's
rules), competes in C$_{60}$ with the Jahn-Teller coupling to
eight fivefold-degenerate vibrational modes of H$_g$
symmetry\cite{Gunnarsson}, which, on the contrary, prefers
low-degeneracy configurations (inverted Hund's rules). If the
vibrational frequencies, ranging from 35 to 196
meV\cite{Gunnarsson}, are larger than the Kondo temperature, one
can neglect retardation effects, which amounts to renormalize $J
\rightarrow J - 3E_{JT}/4$, where $E_{JT}\simeq 169$ meV is the
Jahn-Teller energy gain. Although first principle
calculations\cite{Manini} do predict an inverted
effective-exchange $J - 3E_{JT}/4 \simeq -50$ meV, in what follows
we will consider both $J>0$ and $J<0$ cases.

The impurity is coupled to a bath of conduction electrons, and
both the bath and the hybridization are for simplicity assumed to
be SU(6) and particle-hole ({\em p-h}) invariant. When discussing
the stability of the fixed points, we shall take into account
deviations from this high-symmetry case. The non-interacting
Hamiltonian, ${\cal{H}}_0 = \mathcal{H}_{bath} +
\mathcal{H}_{hyb}$ can always be written as a one-dimensional
chain of non-interacting electrons, $\mathcal{H}_{bath} =
\sum_{\sigma\, m}\, \sum_{i\geq 1} \, t_i \left(c^\dagger_{i\,
m\sigma}\, c^\dagga_{i+1\, m\sigma} + H.c.\right)$, hybridized at
one edge to the impurity, ${\cal{H}}_{hyb} = V\,\sum_{\sigma\,
m}\, \left(d^\dagger_{m\sigma}\, c^\dagga_{1\, m\sigma} +
H.c.\right)$. Here $c^\dagger_{i\, m\sigma}$ and $c^\dagga_{i\,
m\sigma}$ are, respectively, the creation and annihilation
conduction-electron operators at chain site $i$ with spin
$\sigma=\uparrow,\downarrow$ and angular momentum component
$m=-1,0,1$, while $d^\dagger_{m\sigma}$ and $d^\dagga_{m\sigma}$
are the analogous impurity operators.

Although we analyzed by NRG the full AIM, ${\cal{H}}_{AIM} =
{\cal{H}}_{0} + {\cal{H}}_{int}$, the CFT analysis is more
conveniently done for the Kondo model onto which the AIM maps for
large $U$. One finds that ${\cal{H}}_{AIM}\rightarrow
{\cal{H}}_{bath} + {\cal{H}}_{J} + {\cal{H}}_{K}$, where the Kondo
exchange is given by \be {\cal{H}}_{K} = \frac{4V^2}{U}\,
\sum_{l,\lambda,s,\sigma} \, \left\{ c^\dagger_1 \otimes
c^\dagga_1\right\}_{l\lambda,s\sigma}\, \left\{ d^\dagger \otimes
d^\dagga\right\}_{l\lambda,s\sigma}^\dagger. \label{H-Kondo} \ee
$\left\{ d^\dagger \otimes d^\dagga\right\}_{l\lambda,s\sigma} =
\sum \,C^{l\lambda}_{1m\, 1-m'} (-1)^{m'-1} \, C^{s\sigma}_{1/2
\alpha\, 1/2 -\alpha'}$ $(-1)^{\alpha'-1/2} \,
\left(d^\dagger_{m\alpha} d^{\phantom{\dagger}}_{m'\alpha'}
-1/2\,\delta_{mm'}\,\delta_{\alpha\alpha'}\right) $ is the
impurity {\em p-h} operator with angular momentum $l=0,1,2$ and
$z$-component $\lambda$, as well as spin $s=0,1$ and $z$-component
$\sigma$, and analogously for the operator at site $1$ of the
chain. $C^{JJ_z}_{j j_z\,j'j'_z}$ are Clebsch-Jordan coefficients.
Notice that $\left\{ d^\dagger \otimes
d^\dagga\right\}_{00,00}=(n_0-3)/\sqrt{6}$ is zero when $n_0=3$.
Since the U(6) symmetry of ${\cal{H}}_{bath}$ is reduced to charge
U(1), spin SU(2) and orbital O(3), the proper conformal embedding
of the conduction electrons is U(6)$_1$ $\supset$
U(1)$\times$SU(2)$_3\times$SU(2)$_8\times$Z$_3$, where SU(2)$_3$
refers to spin and SU(2)$_8$ to the angular
momentum.\cite{nota-level} The 3-state Potts sector Z$_3$, which
reflects orbital permutational-symmetry, can be interpreted as
emerging out of the charge iso-spin SU(2)$_3$ $\supset$
U(1)$\times$Z$_3$,\cite{CFT} with generators $I_z=\sum_i
(n_i-3)/2$, $ I^+ = \sum_i\, (-1)^i\, \sum_{m=-1}^1\, (-1)^m\,
c^\dagger_{i\, m\uparrow}\, c^\dagger_{i\, -m\downarrow} $, and
$I^-=\left(I^+\right)^\dagger$. They commute with
${\cal{H}}_{bath}$ and also with the Kondo-exchange terms with
$(l,s)=(1,0),(0,1),(2,1)$, while they do not with those having
$(l,s)=(0,0),(2,0),(1,1)$. We recall that the Z$_3$ Potts model
has primary fields $I$, $\sigma$, $Z$ and $\epsilon$, with
dimensions 0, 1/15, 2/3 and 2/5, respectively.\cite{CFT}

Since the AIM is invariant upon $n_0\to 6-n_0$, we only need to
consider $n_0=1,2,3$. For $n_0=1$, ${\cal{H}}_{J}$ is ineffective,
leading at large $U$ to a conventional Kondo-screened SU(6) model.
Conversely, nontrivial behavior may appear for $n_0=2,3$. Here the
Kondo screening, controlled by the Kondo temperature $T_K$, is
hampered by the exchange splitting $J$. While the former takes
advantage from the impurity coherently tunneling among all
available $n_0$-electron configurations, the latter tends to lock
the impurity into a sub-set of states with well defined $S$ and
$L$. When $|J|\sim T_K$, neither prevails and a nontrivial
behavior may emerge.

In the {\em p-h} symmetric case, $n_0=3$, the available impurity
states have quantum numbers $(S,L)=(3/2,0),(1/2,2),(1/2,1)$. For
conventional Hund's rules, $J>0$, the lowest energy state has
$(S,L)=(3/2,0)$, while for $J<0$ $(S,L)=(1/2,1)$ is favored. Let
us first assume a positive and very large $J\gg T_K$. In this case
the impurity effectively behaves as a spin-3/2. If we project
(\ref{H-Kondo}) onto the above impurity configuration, only
$(l,s)=(0,1)$ survives, which is just the spin-exchange
$(8V^2/3U)\, {\bm S}\cdot{\bm S}_1$. The model reduces to a
spin-3/2 impurity coupled to three conduction channels, hence it
describes a perfectly Kondo-screened fixed point which has
Fermi-liquid (FL) behavior. The impurity density of states (DOS),
$\rho(\epsilon)$, is uneffected by ${\cal{H}}_{int}$ at the
chemical potential, $\rho(0)=\rho_0$, meaning a perfect
transmission in the SMT, $G/(2e^2/h) = \rho(0)/\rho_0 = 1$, with
$G$ the zero-bias conductance per orbital.

In the opposite extreme of a strong inverted Hund's rule, $J\ll
-T_K<0$, the impurity locks into the $(S,L)=(1/2,1)$ state. The
Kondo exchange projected onto this subspace contains
$(l,s)=(1,0),(0,1),(2,1)$, hence still commutes with the iso-spin
generators. Thus we expect the asymptotic spectrum to be described
by conformal towers identified by the quantum numbers of the
iso-spin, $I$, spin, $S$, and angular momentum, $L$. To infer the
fixed-point, let us assume a Kondo exchange $J_K\sim V^2/U$ much
larger than the bath bandwidth. In this case we have first to
solve the two-site problem which includes the impurity and the
site 1 of the chain. The lowest energy states are obtained by
coupling into an overall singlet the impurity with a
$(S,L)=(1/2,1)$ electron-configuration. Yet one can form three
states at site 1  with $(S,L)=(1/2,1)$, namely through 1, 3 or 5
electrons. The net result is that the two-site ground state is
threefold degenerate and represents an effective iso-spin-1
impurity. Projecting onto this manifold the hopping term
connecting site 1 to site 2, we get a new Kondo exchange acting
only among iso-spins, namely $\propto (t_1^2/J_K) \, {\bm I}\cdot
{\bm I}_2$. This model describes an iso-spin-1 three-channel Kondo
model which is non-Fermi liquid (NFL). By applying the fusion
hypothesis\cite{A&L}, %%
which is a formal way in CFT to make the impurity dissolve, with
its quantum numbers, into the conduction sea,
we argue that the fixed point is obtained by fusing the $\pi/2$
phase-shifted chain with an iso-spin-1 primary field. The spectrum
obtained in this way agrees with the actual NRG results, see
Table~\ref{3orb-NFL-stable-operators}.\cite{explain-NRG}
\begin{table}
\caption{Left Table: low-energy spectrum at the non-Fermi liquid
stable fixed point for $n_0=3$, in units of the fundamental level
spacing, as predicted by CFT, $E_{CFT}$, and as obtained by NRG,
$E_{NRG}$. Right Table: dimension $x$ of the most relevant
boundary operators. \label{3orb-NFL-stable-operators}}

\begin{center}

\begin{tabular}{|| c | c | c | c | c ||}
\hline\hline
    $I$     &    $S$      &     $L$       & $E_{CFT}$    & $E_{NRG}$ \\
    \hline
    \hline
    1/2    &     0        &     0         &   0   &    0   \\ \hline
     0     &    1/2       &     1         & 1/5   & 0.199   \\ \hline
     1     &    1/2       &     1         & 3/5   & 0.603   \\
    1/2    &     0        &     2         & 3/5   & 0.599   \\
    1/2    &     1        &     1         & 3/5   & 0.599 \\ \hline\hline
\end{tabular}
~~
\begin{tabular}{|| c | c | c | c ||}
    \hline
    \hline
    $I$     &    $S$      &     $L$            & $x$    \\
    \hline
    \hline
     0      &     0       &      0            &  0  \\
    \hline
     1     &      0       &      0            & 2/5 \\
    \hline
     1/2   &     1/2      &      1            & 1/2 \\
     \hline
     0     &      0       &      2            & 3/5 \\
     0     &      1       &      1            & 3/5 \\
%     \hline
%     1/2   &     1/2      &      2            & 9/10 \\
%     1/2   &     3/2      &      0            & 9/10 \\
\hline\hline
  \end{tabular}
\end{center}

\end{table}
The approach to the fixed point is controlled by the leading
irrelevant operator of dimension $7/5$, which is the first
descendant of the iso-spin-1 primary field\cite{A&L} and
represents the residual, actually attractive, interaction among
the conduction electrons once the impurity has been absorbed. This
operator leads to a singular local iso-spin susceptibility,
$\chi_I\sim T^{-1/5}$, whose components are the compressibility
and the pairing susceptibility in the $S=L=0$ Cooper channel. The
fixed point has a residual entropy $S(0)=1/2\, \ln
\left[(\sqrt{5}+1)/(\sqrt{5}-1)\right]$ and an impurity DOS
$\rho(0)=\rho_0/(1+\sqrt{5})$, with $\rho(\epsilon)-\rho(0)\sim
|\epsilon|^{2/5}$.\cite{A&LPRB} This implies that the tunneling
conductance approaches a fractional value at zero temperature,
$G(T=0)/(2e^2/h) = 1/(1+\sqrt{5})$, in a power-law fashion,
$G(T)-G(0)\sim T^{2/5}$. The fixed point is however unstable
towards symmetry-breaking boundary terms which correspond in
Table~\ref{3orb-NFL-stable-operators} to physical operators with
dimensions smaller than one. They include {\em p-h} and gauge
symmetry breaking in the $S=L=0$ Copper-channel, {\sl i.e.} the
$(I,S,L)=(1,0,0)$ operator, as well as quadrupolar distortions and
spin-orbit coupling, the $(I,S,L)=(0,0,2)$ and $(0,1,1)$
operators, respectively.

Since the two stable fixed points for $J\ll -T_K$ and $J\gg T_K$
are essentially different, we expect an unstable fixed point in
between, which we actually find by NRG for a $J_*\sim -T_K$. The
NRG spectrum shows that the iso-spin symmetry is not restored at
this point, unlike at the two stable ones, which forces us to deal
with independent charge U(1) and Potts Z$_3$ sectors. The
conformally invariant boundary conditions of a 3-state Potts model
in two-dimensions are well characterized\cite{Affleck} and include
stable {\sl fixed} boundary conditions, only one of the 3 states
is allowed at the boundary, less stable {\sl mixed} boundary
conditions, two states are allowed, and lastly an unstable {\sl
free} boundary condition. We observe that the Kondo-screened phase
and the non-Fermi liquid one can be identified respectively as the
{\sl fixed} and {\sl mixed} boundary conditions in the Potts
sector. Indeed the non-Fermi liquid fixed point spectrum can also
be obtained fusing the Kondo-screened fixed point with the Z$_3$
primary field $\epsilon$, in agreement with the boundary CFT of
the 3-state Potts model\cite{Affleck}. The unstable fixed point
should then corresponds to the {\sl free} boundary condition. It
has been shown\cite{Affleck} that a proper description of the {\sl
free} boundary condition in the 3-state Potts model requires a
larger set of conformal towers which include the so-called
${\cal{C}}$-disorder fields\cite{Zamo} of dimensions 1/40, 1/8,
21/40 and 13/8. We actually find that the unstable fixed point is
simply obtained by the Kondo-screened one upon fusion with the
same 1/8 primary field that allows to turn the fixed into free
boundary conditions in the 3-state Potts model\cite{Affleck}, see
Table~\ref{3orb-NFL-unstable}.
\begin{table}
\caption{Same as Table~\ref{3orb-NFL-stable-operators} for the unstable fixed point.
This time however the states and operators are labelled by $(Q,S,L,Z_3)$ where
$Q$ refers to the U(1) charge sector and $Z_3$ to the Potts sector.
\label{3orb-NFL-unstable}}

\begin{center}
  \mbox{
    \begin{tabular}{||c|c|c|c|c|c||}
       \hline
       \hline
       $Q$ & $S$ & $L$ & $Z_3$ & $E_{CFT}$ & $E_{NRG}$ \\
       \hline
       1   & 0     & 0   & 1/8   & 0   & 0 \\
       \hline
       0   & 1/2   & 1   & 1/40  & 1/6 & 0.162 \\
       \hline
       1   & 0     & 2   & 1/40  & 1/2 & 0.488 \\
       1   & 1     & 1   & 1/40  & 1/2 & 0.489 \\
       2   & 1/2   & 1   & 1/40  & 1/2 & 0.491 \\
       \hline
       0   & 1/2   & 1   & 21/40 & 2/3 & 0.675 \\
       0   & 1/2   & 2   & 1/8   & 2/3 & 0.643 \\
       0   & 3/2   & 0   & 1/8   & 2/3 & 0.644 \\
       3   & 0     & 0   & 1/8   & 2/3 & 0.648 \\
       \hline
       \hline
    \end{tabular}}
  \hfil
  \mbox{
   \begin{tabular}{||c|c|c|c|c||}
      \hline
      \hline
      $Q$ & $S$ & $L$ & $Z_3$ & $x$ \\
      \hline
      \hline
      0   & 0     & 0   & I           & 0 \\
      \hline
      2   & 0     & 0   & I           & 1/3 \\
      \hline
      1   & 1/2   & 1   & $\sigma$    & 1/2 \\
      \hline
      0   & 0     & 0   & Z           & 2/3 \\
      0   & 0     & 2   & $\sigma$    & 2/3 \\
      0   & 1     & 1   & $\sigma$    & 2/3 \\
      \hline
      1   & 1/2   & 1   & $\epsilon$  & 5/6 \\
      \hline
      \hline
   \end{tabular}}
\end{center}

\end{table}
From the right panel in Table~\ref{3orb-NFL-unstable} it turns out that the
operator which moves away from the fixed point has quantum numbers
$(Q,S,L,Z_3)=(0,0,0,Z)$ and dimension $2/3$. This implies that the
effective energy scale which controls the deviation from the unstable point
behaves as $T_- \sim |J-J_*|^3$.
The most relevant symmetry-breaking operator has
quantum numbers $(Q,S,L,Z_3)=(2,0,0,I)$ and corresponds to the $S=L=0$ Cooper channel.
The corresponding pairing susceptibility diverges at zero temperature $\chi_{SC}\sim T^{-1/3}$.
The next relevant operators of dimension 2/3 are the same quadrupolar distortions and
spin-orbit coupling as for the stable non-Fermi-liquid fixed point.
The residual entropy is $S(0)=1/2\, \ln 3$ and the
impurity DOS $\rho(0)=\rho_0/2$ with $\rho(\epsilon)-\rho(0)\sim -\epsilon^2$, implying
a tunneling conductance $G(0)/(2e^2/h) = 1/2$.

\begin{figure}
\centerline{
\includegraphics[width=5.5cm,height=3cm]{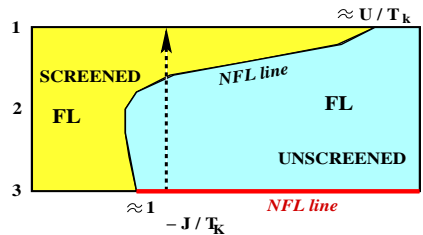}
}
\caption{Qualitative phase diagram of the AIM as function of $n_0$ and $-J/T_K$ for $J<0$.
Both screened and unscreened regions are Fermi liquids (FL). Screened/unscreened
refers to the presence/absence of a Kondo resonance in the DOS.
The critical line separating them is non Fermi liquid (NFL), as well as the line
at $n_0=3$ for $-J/T_K$ above the critical value $\simeq 1$. The arrow
indicates the path along which the DOS in the right panel of
Fig.~\ref{DOS2} is calculated. \label{ph-dia} }
\end{figure}

Unlike the stable NFL fixed point, the unstable one survives {\em
p-h} symmetry breaking. Therefore, although both regimes with
$J\ll -T_K$ and $J\gg T_K$ should have Fermi-liquid behavior away
from {\em p-h} symmetry, an unstable NFL fixed point is still
expected to intrude between them. The NRG analysis confirms this
CFT prediction not only slightly but also far away from {\em p-h}
symmetry. For instance we still observe an unstable fixed point
for $J_*\sim -T_K$ when the average impurity occupancy $n_0=2$. In
this case for $J\ll -T_K$ the impurity locks into a total singlet,
$S=L=0$, which decouples from the conduction sea. For $J\gg T_K$
the impurity instead is in a state with $S=L=1$ which, due to the
$(l,s)=(0,0)$ term in Eq.~(\ref{H-Kondo}), gets fully screened by
four conduction electrons. These two phases turns out to be still
separated by a critical point, which we find can be obtained by
fusing the spectrum of the unscreened fixed point ($J\ll -T_K$)
with the dimension-1/8 primary field of the extended Z$_3$ sector.
The operator content at this unstable fixed point is exactly the
same as in the case of $n_0=3$, thus showing that they are
connected by a whole critical line $J_*(n_0)$. We find that
$J_*\sim -T_K$ for $n_0\in [2,3]$. For $n_0<2$, $|J_*|$ increases
and merges into a mixed-valence critical point $J_*\sim -U$ for
$n_0\to 1$, as sketched qualitatively in Fig.~\ref{ph-dia}.

\begin{figure}[t]
\centerline{
\includegraphics[width=7cm,height=3cm]{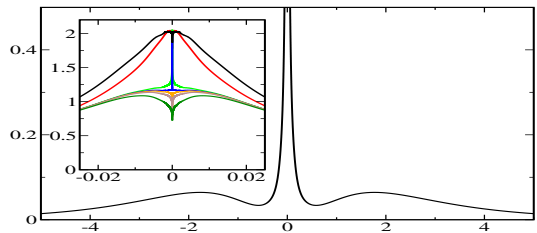}
} \caption{NRG impurity DOS at $n_0=3$ with $U=2$ and $\rho_0=2$
in units of half the bath bandwidth\cite{explain-NRG}. The large
figure shows the DOS on a large scale at $J=0$, where the Hubbard
side-bands and the Kondo resonance are both visible. In the inset
we show the low-energy part across the unstable fixed point. The
curves from the top one downwards correspond to
$J=0.0,-0.01,-0.017,-0.018,-0.02,-0.035,-0.043$. \label{DOS} }
\end{figure}

Let us discuss now the behavior of the impurity DOS in the various phases, which
is directly proportional to the SMT conductance. In Fig.~\ref{DOS} we draw
the DOS across the unstable NFL fixed point for $n_0=3$.
We notice that the DOS displays two distinct features at low energy.
There is a broad resonance which is smooth across the transition.
On top of that, there is a narrow peak in the
Kondo-screened phase which shrinks, disappears at the fixed point,
and transforms in a narrow cusp-like pseudo-gap
within the NFL phase.
We argue\cite{DeLeoPRB} that this dynamical behavior
is controlled by two energy scales. One is the aforementioned scale which
measures the deviation from the fixed point, $T_-\sim |J-J_*|^3$,
and sets the magnitude of the narrow resonance as well as of the
pseudo-gap, $\rho(\epsilon)-\rho(0)\sim \rho(0)\,|\epsilon/T_-|^{2/5}$.
The other is a high energy scale, $T_+\sim |J_*|$, which sets the width of the
broad resonance.

\begin{figure}[t]
\centerline{
\includegraphics[width=7cm,height=3cm]{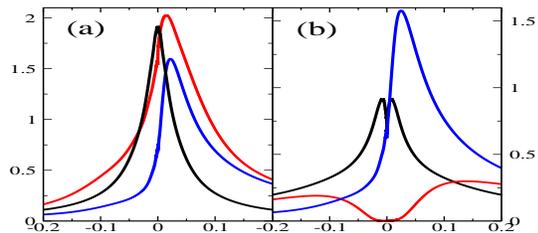}
} \caption{Low-energy impurity DOS at $n_0=1,2,3$. (a) $J=0$. The
peaks of the curves with decreasing $n_0=3,2,1$ gradually move
towards positive energies. (b) $J=-0.075$. $n_0=3$ is the curve
with the cusp at the chemical potential, $n_0=2$ that with the
large pseudo-gap and $n_0=1$ the other one. \label{DOS2} }
\end{figure}

Since it is unlikely that experimentally the system finds itself
near the unstable fixed point, in Fig.~\ref{DOS2} we draw the
low-energy part of the DOS with $n_0=1,2,3$ faraway from the
unstable fixed point. In the figure we show the two cases: 3(a)
$J=0$, standard Kondo effect; 3(b) $J<0$ along a path like the one
drawn in Fig.~\ref{ph-dia}. Unlike the $J=0$ case (also
representative of all positive $J$'s), for $J<0$ the DOS turns out
to be highly sensitive to the doping $n_0$, similarly to what is
observed experimentally\cite{alkali}.

We end by discussing the role of quadrupolar distortions which
lower the icosahedral symmetry of isolated C$_{60}$ and presumably
exist in real SMT devices. In their presence the unstable fixed
point critical region is replaced by a crossover region, more or
less smooth depending on the value of the distortion. Yet the
spectral properties inside the inverted Hund's rule region should
still differ from a conventional perfectly-screened Kondo-like
behavior, particularly when the sensitivity to the doping is
concerned.

In conclusion we have shown that a realistic Anderson impurity model of a
C$_{60}^{n-}$ molecule displays a phase diagram which
includes non-Fermi-liquid phases and critical points
with anomalous properties, most notably
higher sensitivity to doping $n$ and quadrupolar distortions than to magnetic field,
especially for $n>1$.
These results indicate a possibly very interesting
behavior of C$_{60}$ based SMT's.
Provided C$_{60}$ could be endowed with more than one electron in a controllable way,
{\sl e.g.} adsorbed on proper metal substrates\cite{Modesti} or by
alkali doping\cite{alkali}, fractional zero-bias anomalies with power-law
temperature/voltage behavior might eventually show up.
Moreover the enhancement of the pairing fluctuations which emerges out of our analysis
suggests an anomalous behavior when superconducting leads are used,
as for instance an increase of the
zero-bias anomaly below the lead critical-temperature if much smaller than
$T_K$.

In view of recent speculations about the role of
unstable fixed points in single AIM's in the context of
strongly-correlated models on large-coordination lattices\cite{DeLeoPRL},
our results may also have important implications for alkali-doped
fullerenes. We will return to this in a later work.

We acknowledge very helpful discussions with A.O. Gogolin, I. Affleck and E. Tosatti.
This work has been partly supported by MIUR COFIN2004, FIRBRBAU017S8R and
FIRBRBAU01LX5H.


\begin{thebibliography}{30}
\bibitem{explain-NRG} The NRG calculations have been performed with $\Lambda=3$\cite{NRG}
and keeping up to 4500 states per iteration. We implemented the charge U(1), spin SU(2)
and orbital O(3) symmetries.
\end{thebibliography}
\end{document}